\documentstyle[epsf,psfig]{aa}

\begin{document}

\newcommand{\teff}  {${T_{\rm eff}}$}
\newcommand{\mh}    {${\rm [M/H]}$}
\newcommand{\feh}   {${\rm [Fe/H]}$}
\renewcommand{\lg}  {$\log g$}
\newcommand{\mun}    {m$_1$}
\newcommand{\cun}    {c$_1$}
\newcommand{\by}    {b$-$y}
\newcommand{\al}    {\rm et al.}
\newcommand{\eg}    {\em e.g.}
\newcommand{\ie}    {\em i.e.}
\newcommand{\bla}   {\bf bla-bla-bla}
\newcommand\beq{\begin{equation}}
\newcommand\eeq{\end{equation}}
\newcommand\beqa{\begin{eqnarray}}
\newcommand\eeqa{\end{eqnarray}}
\newcommand{\aaa} [2]{A\&A {\bf #1}, #2}
\newcommand{\aas} [2]{A\&A Suppl. {\bf #1}, #2}
\newcommand{\aj}  [2]{AJ {\bf #1}, #2}
\newcommand{\apj} [2]{Ap. J. {\bf #1}, #2}
\newcommand{\apjl}[2]{Ap. J. Letter {\bf #1}, #2}
\newcommand{\apjs}[2]{Ap. J. Suppl. {\bf #1}, #2}
\newcommand{\araa}[2]{A\&AR {\bf #1}, #2}
\newcommand{\pasp}[2]{PASP {\bf #1}, #2}
\newcommand{\mnras}[2]{MNRAS {\bf #1}, #2}

\thesaurus{08 (08.06.3;    
               08.02.1;    
               08.19.1)    
          }

 \title{Metallicity-dependent  effective temperature determination  for
 eclipsing binaries from synthetic $uvby$ Str{\"o}mgren photometry }

\author{E.   Lastennet$^{\star ,}$   \inst{1,2}, T.    Lejeune    \inst{1}, 
P. Westera \inst{1} \and R. Buser \inst{1}}

\institute{Astronomisches Institut der Universit\"at Basel,
           Venusstr. 7, CH--4102 Binningen, Switzerland
     \and  UMR CNRS 7550, Observatoire Astronomique, 
            11, rue de l'Universit\'e, 67000 Strasbourg, France}

\offprints{E.Lastennet@qmw.ac.uk \\
$\star$ present address: Astronomy Unit, Queen Mary and Westfield College, 
Mile End Road, London E1 4NS, UK}

\date{Received September 10, 1998 / Accepted November 5, 1998 }

\authorrunning{E. Lastennet {\al}}
\titlerunning{Metallicity-dependent effective temperature determination}

\maketitle

\begin{abstract}

Str{\"o}mgren synthetic photometry from an empirically calibrated grid
of  stellar atmosphere  models has been  used  to derive the effective
temperature  of  each component  of double   lined spectroscopic (SB2)
eclipsing binaries.  For  this purpose, we  have selected a sub-sample
of 20 SB2s for  which (\by), m$_1$,  and c$_1$ individual indices  are
available.   This new determination of  effective temperature has been
performed in a  homogeneous way for all these  stars. As the effective
temperature determination  is related to  the  assumed metallicity, we
explore   simultaneous solutions  in  the ({\teff},{\feh})-plane   and
present  our results  as  confidence  regions  computed to  match  the
observed values  of surface gravity,  (\by),  m$_1$, and c$_1$, taking
into account  interstellar reddening.   These confidence regions  show
that previous estimates of {\teff} are often  too optimistic, and that
{\feh}  should not be  neglected  in such determinations.  Comparisons
with Ribas {\al} (1998)  using Hipparcos parallaxes are also presented
for 8 binaries of our working sample,  showing good agreement with the
most reliable parallaxes. This point gives a significant weight to the
validity of the BaSeL models for synthetic photometry applications.


\end{abstract}

\section{Introduction}

Fundamental  stellar  parameters such  as   masses  and radii  of well
detached  double-lined  spectroscopic   eclipsing  binaries   can   be
determined very  accurately (Andersen 1991).  Therefore, from accurate
(1-2\%)  stellar mass and radius determinations  of  such objects, one
can compute surface gravities  to very high and otherwise inaccessible
confidence levels.  Indeed, Henry \&  Mc Carthy (1993) have  discussed
the data available for visual binaries of solar mass  and below.  Only
$\alpha$  Cen B (G2V,0.90   M$_{\odot}$)  has a  mass  known  with  an
accuracy comparable to that  for  favorable eclipsing binaries.   This
point shows the  importance of  choosing such  double-lined  eclipsing
binaries  in  order  to obtain surface    gravities  with the  highest
possible accuracy.  Moreover, these binaries are  of great interest to
perform  accurate   tests of  stellar   evolutionary models  (see e.g.
Lastennet {\al} 1996, Pols {\al} 1997, Lastennet \& Valls-Gabaud 1998)
used to derive cluster ages.    The knowledge of all possible  stellar
parameters for such single stars is the  basis of the modelling of the
global physical properties and evolution of star clusters or galaxies.
Nevertheless, while masses  and    radii are accurately    known,  the
effective temperatures --  and consequently, the luminosities of these
stars  --   strongly depend  upon   the  calibration  used   to relate
photometric indices  with   {\teff}.  As a matter  of   fact, for such
binaries the temperatures given  in  the literature come from  various
calibration   procedures  and  are    indeed   highly   inhomogeneous.
Furthermore,  due to the  lack of empirical  calibrations at different
metallicities,  solar metallicity   is often  assumed  for photometric
estimations of     {\teff}.  In  this regard,    synthetic  photometry
performed from large grids of stellar atmosphere models, calibrated in
{\teff}, {\lg}, and {\feh}, provides a powerful tool of investigation.
In   this  paper, we explore    simultaneous solutions of  {\teff} and
{\feh},   and  we  address   the    question  of  the reliability   of
metallicity-independent effective temperature determinations.

We have selected 20 binary systems (40 stars) for which we have $uvby$
Str{\"o}mgren photometry with estimated errors (see Table 1).
For this sample,  previous estimates of  effective temperature are not
homogeneous,  originating  from  various calibrations  established for
different  spectral-type domains: Morton   \& Adams (1968), Relyea  \&
Kurucz (1978),  Osmer  \&  Peterson (1974), Grosb{\o}l  (1978),  Davis \&
Shobbrook (1977), Popper (1980),  Moon \& Dworetsky (1985), Saxner \&
Hammarb\"ack (1985), Jakobsen (1986),  Magain (1987), Napiwotzki {\al}
(1993), and Edvardsson {\al}  (1993).  Moreover, all these studies are
of   course historically not  fully  independent.  As  an example, the
{\teff} of Moon \& Dworetsky   (1985) is estimated using the  {\teff},
(B$-$V)$_0$    calibration of  Hayes (1978)   and   the {\teff}, c$_0$
calibration of Davis  \&  Shobbrook (1977).  However, this   does not
mean  that  these  calibrations   allow     to derive  very    similar
temperatures.  As highlighted by Andersen \& Clausen (1989) concerning
the O-type  components of EM Carinae,  the temperature  calibration of
Davis \& Shobbrook (1977), Jakobsen  (1986), and Popper (1980) do not
agree  particularly well.    A  similar  comparison  of these    three
calibrations  made  by Clausen \&   Gim\'enez  (1991) with the massive
B-type  components of CW Cephei  leads to $\Delta${\teff}$\sim$5500K !
Thus,  a  new  {\em  and}   homogeneous  determination  of   effective
temperature is of primordial interest for such well-known objects.  In
order to re-derive in a homogeneous way the {\teff} of these stars, we
have used the {\em Ba}sel {\em  S}tellar {\em L}ibrary (hereafter {\em
``BaSeL''}) which provides empirically calibrated model spectra over a
large range of stellar parameters (Lejeune {\al} 1997, 1998a).

In  Section \ref{sect:modellib}, we will   describe the models used to
perform our calculation of {\teff} from $uvby$ Str{\"o}mgren photometry.
Sect.  \ref{sect:effectemp} will be  devoted to the description of the
method and the presentation of the results.

\section{Model colours}
\label{sect:modellib}

The BaSeL models cover a large range of fundamental parameters: 2000 K
$\leq$ {\teff} $\leq$  50,000 K, $-$1.02  $\leq$ {\lg} $\leq$ 5.5, and
$-$5.0 $\leq$ {\mh}   $\leq$ +1.0.  This library  combines theoretical
stellar energy distributions which are based on several original grids
of blanketed model atmospheres, and which have  been corrected in such
a way as to provide synthetic colours consistent with extant empirical
calibrations  at all wavelengths from the   near-UV through the far-IR
(see Lejeune {\al}  1997, 1998a).  For our  purpose, we have  used the
new version of the BaSeL models  for which the correction procedure of
the theoretical  spectra  has been  extended   to higher  temperatures
({\teff} $\ge$ 12,000   K), using the  {\teff}--(B$-$V) calibration of
Flower (1996),  and   to shorter wavelengths   (Lejeune  {\al} 1998b).
Because  the    correction   procedure  implies  modulations   of  the
(pseudo-)continuum   which    are  smooth    between the   calibration
wavelengths,   the   final  grid   provides   colour-calibrated   flux
distributions   (9.1 $\leq  \lambda \leq$  160,000  nm,   with a  mean
resolution of 1 $\sim$ 2 nm from the UV to the visible) which are also
suitable for  calculating  medium-band  synthetic  photometry, such as
Str\"omgren colours.   Thus,  synthetic  Str{\"o}mgren photometry  was
performed using the  passband response functions ($u,  v, b, y$) given
in   Schmidt-Kaler (1982).   Theoretical  (u$-$b),  (\by),  {\mun} $=$
(v$-$b)$-$(\by), and   {\cun} $=$ (u$-$v)$-$(v$-$b) indices  have been
computed, where the zero-points  were defined by matching the observed
colours   (u$-$b $=$ 1.411,   \by    $=$  0.004, {\mun} $=$ 0.157,
{\cun}  $=$   1.089; Hauck \& Mermilliod   1980)   of Vega  with those
predicted by  the  corresponding Kurucz (1991)  model  for {\teff} $=$
9400 K, {\lg} $=$ 3.90, {\mh} $=$ $-$0.50.

\section{Effective temperature determination}
\label{sect:effectemp}

\subsection{Str{\"o}mgren data} 

Among  the approximately   sixty  SB2 systems   gathered in  Lastennet
(1998), only 20 have   both individual $uvby$ Str{\"o}mgren  photometric
indices and uncertainties for each component.  Uncertainties are a key
point  in the calculation  presented  later (Sect. \ref{sect:method}).
The  photometry used for  the 20  systems of  our working  sample (see
Table 1)
is from the recent Table 5 of  Jordi {\al} (1997),  who have taken the
individual indices directly  from the literature  but  have also added
their own results for three systems (YZ Cas, WX Cep, and IQ Per).

\begin{table*}[h]
\begin{center}
	\caption[]{Str{\"o}mgren photometry for the sample (after Table 5 of Jordi {\al} 1997). 
                   Some useful notes about reddening are given in the last column.}
\begin{flushleft}
\begin{tabular}{lccccll}
\hline\noalign{\smallskip}
System &  (b$-$y)   &   m$_1$   &  c$_1$ &   {\lg}  & E(b$-$y)$^\dag$ &  E(b$-$y) \\
\noalign{\smallskip}
\hline\noalign{\smallskip}
BW Aqr      &  0.345$\pm$0.015  & 0.15$\pm$0.03    & 0.45$\pm$0.03    & 3.981$\pm$0.020 & 0.03    & 0.03           $^{(a)}$ \\ 
            &  0.325$\pm$0.015  & 0.16$\pm$0.03    & 0.45$\pm$  0.03  & 4.075$\pm$0.022 & 0.03    & 0.03           $^{(a)}$  \\  
AR Aur      & -0.043$\pm$0.010  & 0.142$\pm$0.012  & 0.857$\pm$ 0.015 & 4.331$\pm$0.025 & 0.      & 0.             $^{(z)}$  \\  
            & -0.021$\pm$0.010  & 0.162$\pm$0.012  & 0.892$\pm$0.015  & 4.280$\pm$0.025 & 0.      & 0.             $^{(z)}$  \\  
$\beta$ Aur & -0.003$\pm$0.026  & 0.162$\pm$0.053  & 1.124$\pm$0.057  & 3.930$\pm$0.010 & 0.      & 0.             $^{(z)}$  \\  
            &  0.005$\pm$0.026  & 0.206$\pm$0.053  & 1.121$\pm$0.057  & 3.962$\pm$0.010 & 0.      & 0.             $^{(z)}$  \\  
GZ Cma      &  0.077$\pm$0.010  & 0.193$\pm$0.020  & 1.066$\pm$0.025  & 3.989$\pm$0.012 & 0.047   & 0.047$\pm$0.02 $^{(b)}$   \\  
            &  0.091$\pm$0.010  & 0.216$\pm$0.020  & 1.002$\pm$0.025  & 4.083$\pm$0.016 & 0.047   & 0.047$\pm$0.02 $^{(b)}$   \\  
EM Car      &  0.310$\pm$0.010  & -0.038$\pm$0.010 & -0.089$\pm$0.010 & 3.857$\pm$0.017 & 0.44    & 0.44           $^{(c)}$  \\  
            &  0.310$\pm$0.010  & -0.047$\pm$0.010 & -0.076$\pm$0.010 & 3.928$\pm$0.016 & 0.44    & 0.44           $^{(c)}$  \\  
YZ Cas      &  0.004$\pm$0.006  & 0.186$\pm$0.009  & 1.106$\pm$0.011  & 3.995$\pm$0.011 & 0.      & 0.             $^{(z)}$ \\  
            &  0.248$\pm$0.081  & 0.196$\pm$0.166  & 0.309$\pm$0.238  & 4.309$\pm$0.010 & 0.      & 0.             $^{(z)}$ \\  
WX Cep      &  0.330$\pm$0.007  & 0.105$\pm$0.012  & 1.182$\pm$0.023  & 3.640$\pm$0.011 & 0.3     & 0.             $^{(z)}$  \\  
            &  0.271$\pm$0.022  & 0.080$\pm$0.036  & 1.190$\pm$0.060  & 3.939$\pm$0.011 & 0.3     & 0.             $^{(z)}$ \\  
CW Cep      &  0.333$\pm$0.010  & -0.071$\pm$0.015 & 0.037$\pm$0.015  & 4.059$\pm$0.024 & 0.46    & 0.46           $^{(d)}$ \\  
            &  0.339$\pm$0.010  & -0.064$\pm$0.015 & 0.045$\pm$0.015  & 4.092$\pm$0.024 & 0.46    & 0.46           $^{(d)}$ \\  
RZ Cha      &  0.314$\pm$0.016  & 0.149$\pm$0.027  & 0.480$\pm$0.027  & 3.909$\pm$0.009 & 0.003   & 0.             $^{(z)}$ \\  
            &  0.304$\pm$0.017  & 0.165$\pm$0.029  & 0.468$\pm$0.029  & 3.907$\pm$0.010 & 0.003   & 0.             $^{(z)}$ \\  
KW Hya      &  0.105$\pm$0.005  & 0.243$\pm$0.007  & 0.919$\pm$0.005  & 4.079$\pm$0.013 & 0.01    & 0.             $^{(e),(z)}$ \\  
            &  0.244$\pm$0.011  & 0.210$\pm$0.007  & 0.490$\pm$0.047  & 4.270$\pm$0.010 & 0.01    & 0.             $^{(e),(z)}$ \\  
GG Lup      & -0.049$\pm$0.007  & 0.097$\pm$0.011  & 0.450$\pm$0.012  & 4.301$\pm$0.012 & 0.020   & 0.020          $^{(f)}$  \\  
            & -0.019$\pm$0.019  & 0.141$\pm$0.032  & 0.811$\pm$0.036  & 4.364$\pm$0.010 & 0.020   & 0.020          $^{(f)}$  \\  
TZ Men      & -0.025$\pm$0.007  & 0.140$\pm$0.010  & 0.941$\pm$0.010  & 4.225$\pm$0.011 & 0.      & 0.             $^{(z)}$ \\  
            &  0.185$\pm$0.007  & 0.176$\pm$0.015  & 0.689$\pm$0.015  & 4.303$\pm$0.009 & 0.      & 0.             $^{(z)}$ \\  
V451 Oph    &  0.084$\pm$0.010  & 0.083$\pm$0.020  & 0.940$\pm$0.020  & 4.038$\pm$0.015 & 0.115   & 0.115          $^{(g)}$  \\  
            &  0.103$\pm$0.010  & 0.109$\pm$0.020  & 0.992$\pm$0.020  & 4.196$\pm$0.015 & 0.115   & 0.115          $^{(g)}$   \\  
V1031 Ori   &  0.10$\pm$0.01    & 0.17$\pm$0.02    & 1.13$\pm$0.03    & 3.560$\pm$0.008 & 0.05    & 0.             $^{(h)}$  \\  
            &  0.05$\pm$0.01    & 0.16$\pm$0.02    & 1.13$\pm$0.03    & 3.850$\pm$0.019 & 0.05    & 0.             $^{(h)}$  \\  
IQ Per      &  0.056$\pm$0.004  & 0.079$\pm$0.005  & 0.635$\pm$0.011  & 4.208$\pm$0.019 & 0.11    & 0.10$\pm$0.01  $^{(i)}$ \\  
            &  0.165$\pm$0.049  & 0.089$\pm$0.103  & 0.819$\pm$0.186  & 4.323$\pm$0.013 & 0.11    & 0.10$\pm$0.01  $^{(i)}$  \\  
AI Phe      &  0.528$\pm$0.010  & 0.308$\pm$0.010  & 0.379$\pm$0.010  & 3.593$\pm$0.003 & 0.015   & 0.015$\pm$0.02 $^{(j)}$  \\  
            &  0.316$\pm$0.010  & 0.172$\pm$0.010  & 0.421$\pm$0.010  & 4.021$\pm$0.004 & 0.015   & 0.015$\pm$0.02 $^{(j)}$  \\  
$\zeta$ Phe & -0.07$\pm$0.02    & 0.13$\pm$0.03    & 0.49$\pm$0.03    & 4.122$\pm$0.009 & 0.      & 0.             $^{(z)}$ \\  
            & -0.01$\pm$0.02    & 0.11$\pm$0.03    & 0.77$\pm$0.03    & 4.309$\pm$0.012 & 0.      & 0.             $^{(z)}$ \\  
PV Pup      &  0.201$\pm$0.024  & 0.171$\pm$0.041  & 0.628$\pm$0.041  & 4.257$\pm$0.010 & 0.06    & 0.             $^{(z)}$  \\  
            &  0.201$\pm$0.025  & 0.159$\pm$0.043  & 0.640$\pm$0.043  & 4.278$\pm$0.011 & 0.06    & 0.             $^{(z)}$  \\  
VV Pyx      &  0.016$\pm$0.006  & 0.156$\pm$0.010  & 1.028$\pm$0.010  & 4.089$\pm$0.009 & 0.016   & 0.016          $^{(k)}$  \\  
            &  0.016$\pm$0.006  & 0.156$\pm$0.010  & 1.028$\pm$0.010  & 4.088$\pm$0.009 & 0.016   & 0.016          $^{(k)}$  \\  
DM Vir      &  0.317$\pm$0.007  & 0.171$\pm$0.010  & 0.480$\pm$0.012  & 4.108$\pm$0.009 & 0.017   & 0.017          $^{(l)}$  \\  
            &  0.317$\pm$0.007  & 0.171$\pm$0.010  & 0.480$\pm$0.012  & 4.106$\pm$0.009 & 0.017   & 0.017          $^{(l)}$  \\  
\noalign{\smallskip}\hline
\end{tabular}
\end{flushleft}
\end{center}
\small $^\dag$ this  work   (cf. Sect  \ref{sect:red})  \\  $^{(a)}$  Clausen
(1991); $^{(b)}$  Popper  {\al} (1985);  $^{(c)}$ Andersen \&  Clausen
(1989); $^{(d)}$ Clausen  \& Gim\'enez (1991);  $^{(e)}$ Our  value is
consistent    with  E(b$-$y)$=$0.009$\pm$0.008  for A-stars (Crawford,
1979); $^{(f)}$ Andersen  {\al}  (1993) using  the (b$-$y)$_0$$-$c$_0$
relation of Crawford (1978);  $^{(g)}$ Clausen {\al} (1986) determined
the reddening from the  [u$-$b]$-$(b$-$y)$_0$ relation  for early-type
stars     of    Str{\"o}mgren  \&   Olsen    (unpublished)     and the
c$_0$$-$(b$-$y)$_0$  relation  of Crawford   (1973), which give nearly
identical results; $^{(h)}$ Andersen {\al} (1990) used E(b$-$y)$=$ 0.0
but   quote    E(b$-$y)$=$0.025    as a    possible  value;   $^{(i)}$
E(B$-$V)$=$0.14$\pm$0.01   (Lacy        \& Frueh    1985);    $^{(j)}$
E(B$-$V)$=$0.02$\pm$0.02 (Hrivnak \&  Milone 1984);  $^{(k)}$ Andersen
{\al} (1984)   using the calibrations   of Grosb{\o}l (1978); $^{(l)}$
Moon \&   Dworetsky  (1985); $^{(z)}$ At the   best  of our knowledge,
systems  for which   interstellar  reddening  has  been  neglected  or
considered as   insignificant in  the  literature.\\ Note:   we assume
E(b$-$y)$=$0.73$\times$E(B$-$V) after Crawford (1975). \normalsize\\
\end{table*}
 
 \subsection{Methodology}
 \label{sect:method}

To compute synthetic colours from the BaSeL  models, we need effective
temperature   (\teff),   surface   gravity ({\lg}),   and  metallicity
({\feh}).   Consequently, given the   observed colours (namely, b$-$y,
m$_1$,  and c$_1$), we are able  to derive {\teff},  {\lg}, and {\feh}
from a comparison with model colours.  As the surface gravities can be
derived very accurately from the masses and radii  of the stars in our
working  sample, only two    physical parameters have  to  be  derived
({\teff} and {\feh}).

This has been done by minimizing the $\chi^2$-functional, defined as

\beqa 
 \chi^2 (T_{\rm eff}, [Fe/H])  = \sum_{i=1}^{n} \left[ \left(\frac{\rm
 colour(i)_{\rm   syn} -  colour(i)}{\sigma(\rm   colour(i))}\right)^2
 \right],
\eeqa 

where  $n$ is the  number   of comparison data, colour(1)=  (\by)$_0$,
colour(2)=  m$_0$,  and  colour(3) =  c$_0$.    The  best $\chi^2$  is
obtained when the synthetic colour, colour(i)$_{\rm syn}$, is equal to
the observed one.

Reddening has   been  taken into  account   following Crawford (1975):
(b$-$y)$_0$  = (b$-$y)  $-$ E(b$-$y),  m$_0$  = m$_1$  + 0.3  $\times$
E(b$-$y), c$_0$ = c$_1$ $-$ 0.2 $\times$  E(b$-$y), in order to derive
the    intrinsic colours  from      the observed  ones.   With   $n=3$
observational data (\by,    m$_1$, c$_1$) and  $p=2$  free  parameters
({\teff} and {\feh}), we  expect to find a $\chi^2$-distribution  with
$q=n-p=1$  degree  of freedom.    Finding  the central minimum value
$\chi^{2}_{\rm min}$, we   form  the $\chi^2$-grid in   the  ({\teff},
{\feh})-plane and compute  the boundaries corresponding to 1 $\sigma$,
2 $\sigma$, and 3 $\sigma$ respectively.   As our sample contains only
stars belonging to the Galactic disk,
we  have  explored a  restricted range of  metallicity,  $-$1.0 $\leq$
{\feh} $\leq$ +0.5.

\begin{figure}[!hb]
\centerline{\psfig{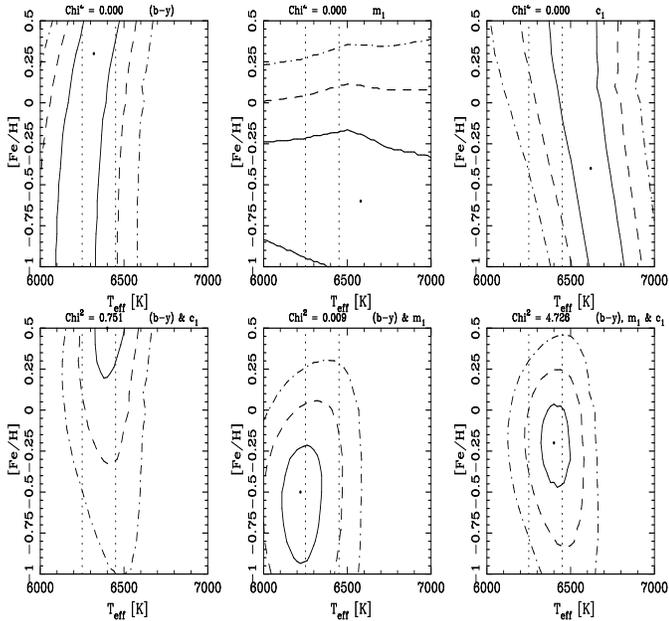}}

	\caption{Simultaneous solutions of {\teff}  and {\feh} for  BW
	Aqr A (assuming  {\lg} = 3.981):  matching (b$-$y) ({\it upper
	left}),  m$_1$  ({\it  upper   central}),  c$_1$  ({\it  upper
	right}), (b$-$y), and c$_1$   ({\it lower left}), (b$-$y),  and
	m$_1$ ({\it  lower central}), (b$-$y),  m$_1$, and c$_1$ ({\it
	lower  right}).  Best fit ({\it black dot}) and 1-$\sigma$ ({\it
	solid    line}),    2-$\sigma$   ({\it    dashed   line}), and
	3-$\sigma$({\it  dot-dashed  line}) confidence levels are also
	shown.  Previous estimates of {\teff}  from Clausen (1991) are
	indicated as vertical dotted lines in all panels.}
\label{fig:example} 
\end{figure}
Figure  \ref{fig:example}  illustrates  the  different  steps  of  the
method, here for BW Aqr A.  All possible combinations of observational
data (as indicated on  the  top of each   panel) are  explored,  hence
varying the number of degrees of freedom  for minimizing the $\chi^2$.
The top   panels show the results obtained   for matching uniquely one
colour index ({\by}, {\mun}, or {\cun}).  In these cases, $q = n - p
= -1$, which simply  means that it is impossible  to fix  both {\teff}
and    {\feh}  with only one   observational    quantity, as indeed is
illustrated by the three  top  panels of Fig.\ref{fig:example}.   From
(\by) only, the effective   temperature boundaries appear to  be  very
similar across the whole metallicity range, highlighting the fact that
this  index is traditionally used  to  derive {\teff}.  Alternatively,
the  {\mun} index only provides  constraints on the metallicity of the
star.  Used together (lower central  panel), these two indices outline
restricted ``islands'' of  solutions  in the ({\teff},  {\feh})-plane,
and hence offer a good combination to  estimate these parameters.  The
{\cun}  index  has been  originally  designed to estimate  the surface
gravity, but it also appears to be a  good indicator of temperature in
the parameter range explored for  BW Aqr A (upper  right panel).  On
the   lower   right panel, {\em    all}   the available  observational
information ({\by}, {\mun}, {\cun}, and {\lg}) can  be exploited.  The
range of {\teff} values  that we then derive  for BW Aqr A agree  well
with  previous estimates (as indicated  by the vertical dotted lines),
and the same is true for its metallicity, which is compatible with the
Galactic disk stars.  Finally, in order to take  full advantage of all
the  observational information available for  the stars in our sample,
we choose to estimate {\teff} and {\feh}  from a $\chi^2$ minimization
performed on the three colour indices.

\subsection{Surface gravity accuracy and influence of reddening}

\subsubsection{Surface gravity}

As our  results depend  not only  on the accuracy  of  the photometric
data,  but  also on  that  of the   surface  gravity determination, we
analysed the effect of a variation of {\lg} upon the predicted {\teff}
and {\feh} values.  We investigated  this ``{\lg} effect'' for the  AR
Aur  system, for  which  the known   value  of {\lg} has   the largest
uncertainties in our working sample: for instance, the surface gravity
of   the coolest   component of   AR Aur  (AR  Aur B)     is {\lg} $=$
4.280$\pm$0.025.

\begin{figure}[ht]
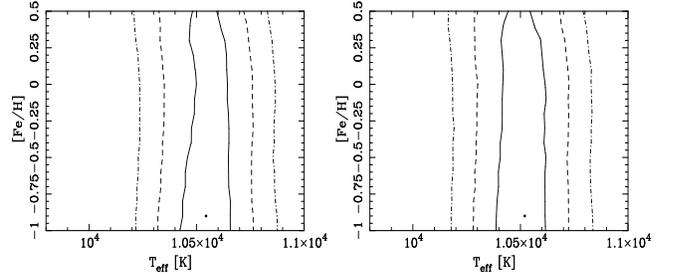

 \centerline{
 \psfig{file=8144.01.f2,width=4.2cm,height=3.6truecm,rheight=3.9truecm,angle=-90.}
 \psfig{file=8144.02.f2,width=4.2cm,height=3.6truecm,rheight=3.9truecm,angle=-90.}
 } 
	\caption{Influence  of  {\lg} on the  simultaneous solution of
	{\teff} and  {\feh} for AR  Aur B.  Two different {\lg} values
	are considered: {\lg} $=$ 4.255  ({\it left panel}), {\lg} $=$
	4.305 ({\it right panel}).  These  values are 0.025 dex higher
	or  lower than   the true  {\lg} (4.280).}   \label{fig:araur}
	\end{figure}

For {\lg}  $=$  4.280, the central  {\teff} value  predicted is  about
10,500K.     If we  consider   {\lg}$-$0.025    dex  (left panel    in
Fig.  \ref{fig:araur}) or   {\lg}$+$0.025  dex (right   panel  in Fig.
\ref{fig:araur}), neither the central {\teff} value nor the pattern of
contours change significantly.

This example shows that    our results for ({\teff},{\feh})  will  not
change  due to variations of  surface gravity within the errors listed
in Table 1.

\subsubsection{Interstellar reddening}
\label{sect:red}

Interstellar reddening is of prime importance for the determination of
both {\teff}  and  {\feh}.  A great   deal of attention  was therefore
devoted  to the E(b$-$y) values  available in the literature, for each
star  of our   sample.  We   explore different  reddening  values  (as
described in Sect.  \ref{sect:method}), and we compare their resulting
$\chi^2$-scores. For the following  systems, we adopted the  published
values, in perfect agreement with our results: BW Aqr, AR Aur, $\beta$
Aur, GZ Cma, EM Car, CW Cep, GG Lup, TZ Men, V451 Oph, AI Phe, $\zeta$
Phe, VV Pyx, and DM Vir.  As we did  not find any indication about the
interstellar reddening  of YZ Cas,  we kept E(b$-$y) $=$  0 as a quite
reasonable hypothesis.  We have neither found any data on interstellar
reddening  for   the WX Cephei    system.  But the   hypothesis  of no
significant reddening  for   WX Cep is  ruled   out by the   very high
$\chi^2$-value obtained  in  reproducing simultaneously the quadruplet
(b$-$y, m$_1$,  c$_1$, \lg) of the observed  data.  From the different
reddening values  explored in Table 2, we  find that E(b$-$y) $=$ 0.32
for WX  Cep A  and  E(b$-$y) $=$  0.28 for WX  Cep B  provide the best
solutions.
\begin{table}[h]
\begin{center}
	\caption[]{Influence of  reddening  on  the $\chi^2$  of   the
	components of the WX Cephei system.  Best $\chi^2$ are in bold
	characters. The hypothesis  of  no reddening   is definitively
	ruled out.}
\begin{flushleft}
\begin{center}
\begin{tabular}{crr}
\hline\noalign{\smallskip}
           & WX Cep A        &  WX Cep B        \\
 E(b$-$y)  & $\chi^2$-values & $\chi^2$-values  \\
\noalign{\smallskip}
\hline\noalign{\smallskip}
0.00 & 912.890     & 115.290      \\
0.26 & 29.039      & 0.933       \\
0.28 & 10.974      & {\bf 0.682} \\
0.30 & 4.549       & 1.210       \\
0.32 & {\bf 4.009} & 2.651       \\
0.34 & 4.322       & 4.972       \\
\noalign{\smallskip}
\hline
\end{tabular}
\end{center}
\end{flushleft}
\end{center}
\label{tab:WXCEP}
\end{table}

\begin{figure}[ht]
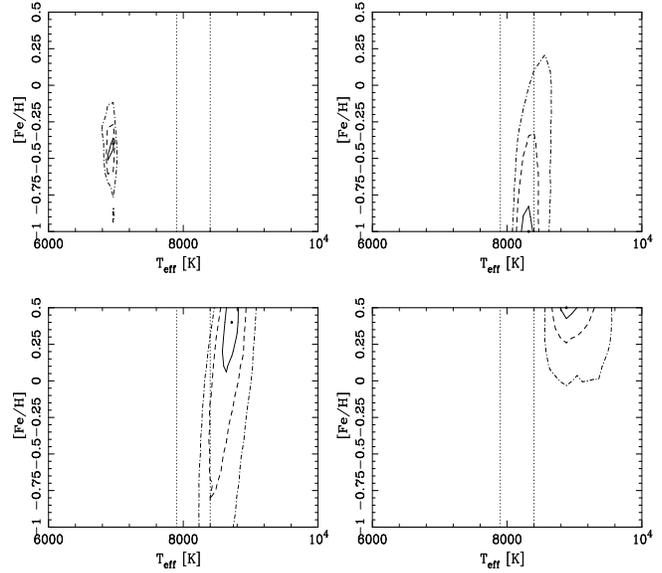

\centerline{\psfig{file=8144.01.f3,width=4.2cm,height=3.6truecm,rheight=3.9truecm,angle=-90.}
            \psfig{file=8144.02.f3,width=4.2cm,height=3.6truecm,rheight=3.9truecm,angle=-90.}}
\centerline{\psfig{file=8144.03.f3,width=4.2cm,height=3.6truecm,rheight=3.9truecm,angle=-90.}
            \psfig{file=8144.04.f3,width=4.2cm,height=3.6truecm,rheight=3.9truecm,angle=-90.}}
	\caption{Influence  of reddening on the simultaneous solutions
	of  {\teff}  and  {\feh} for   WX Cep A.   Different reddening
	values are considered: E(b$-$y) $=$  0.00 ({\it upper  left}),
	E(b$-$y) $=$ 0.30 ({\it upper right}), E(b$-$y) $=$ 0.32 ({\it
	lower   left}), and E(b$-$y)  $=$   0.34 ({\it lower  right}).
	Previous determination  of  {\teff} from Popper (1987)  (using
	Popper's 1980  calibrations) is also  shown for comparison ({\it
	vertical dotted lines}).}  \label{fig:wxcep} \end{figure}

The  influence   of  reddening variations   is illustrated   in Figure
\ref{fig:wxcep}. While  for  the system, WX Cep   AB, an average value
E(\by) $=$ 0.30  appears justified from the  results of Table 2 -- and
will  indeed be adopted in  the remainder of this   paper --, Figure 3
shows how  for  the  individual component,  WX  Cep  A, small  changes
$\Delta$E(\by)$= \pm$0.02 away from  its own optimum value  E(\by) $=$
0.32 induce significant  changes   in the possible solutions    of the
({\teff}, {\feh})-couples.  In   particular, going to E(\by)  $=$ 0.30
(upper right panel) implies a dramatic jump in predicted {\feh} from a
(plausible)  metal-normal (lower  left)    to a (rather unlikely    ?) 
metal-poor composition at or near  the lower limit  ({\feh} $= -1$) of
the exploration range.

For  the other four systems  for which interstellar reddening has also
been  previously neglected in the  literature, we found small, but not
significant, E(b$-$y) values: RZ Cha (0.003), KW Hya (0.01), V1031 Ori
(0.05), and PV Pup (0.06).  

E(b$-$y)   $=$ 0.11  has been   adopted for IQ   Per   by  comparing
different  $\chi^2_{\rm min}$ solutions.   This  value is consistent
with E(b$-$y) $=$ 0.10$\pm$0.01, estimated from the published value of
E(B$-$V) $=$ 0.14$\pm$0.01 (Lacy \& Frueh 1985), assuming E(b$-$y) $=$
0.73$\times$E(B$-$V) after  Crawford (1975).  Adopted reddening values
for stars of our sample are listed in Table 1.

\subsection{General results and discussion}

\begin{figure*}[ht]
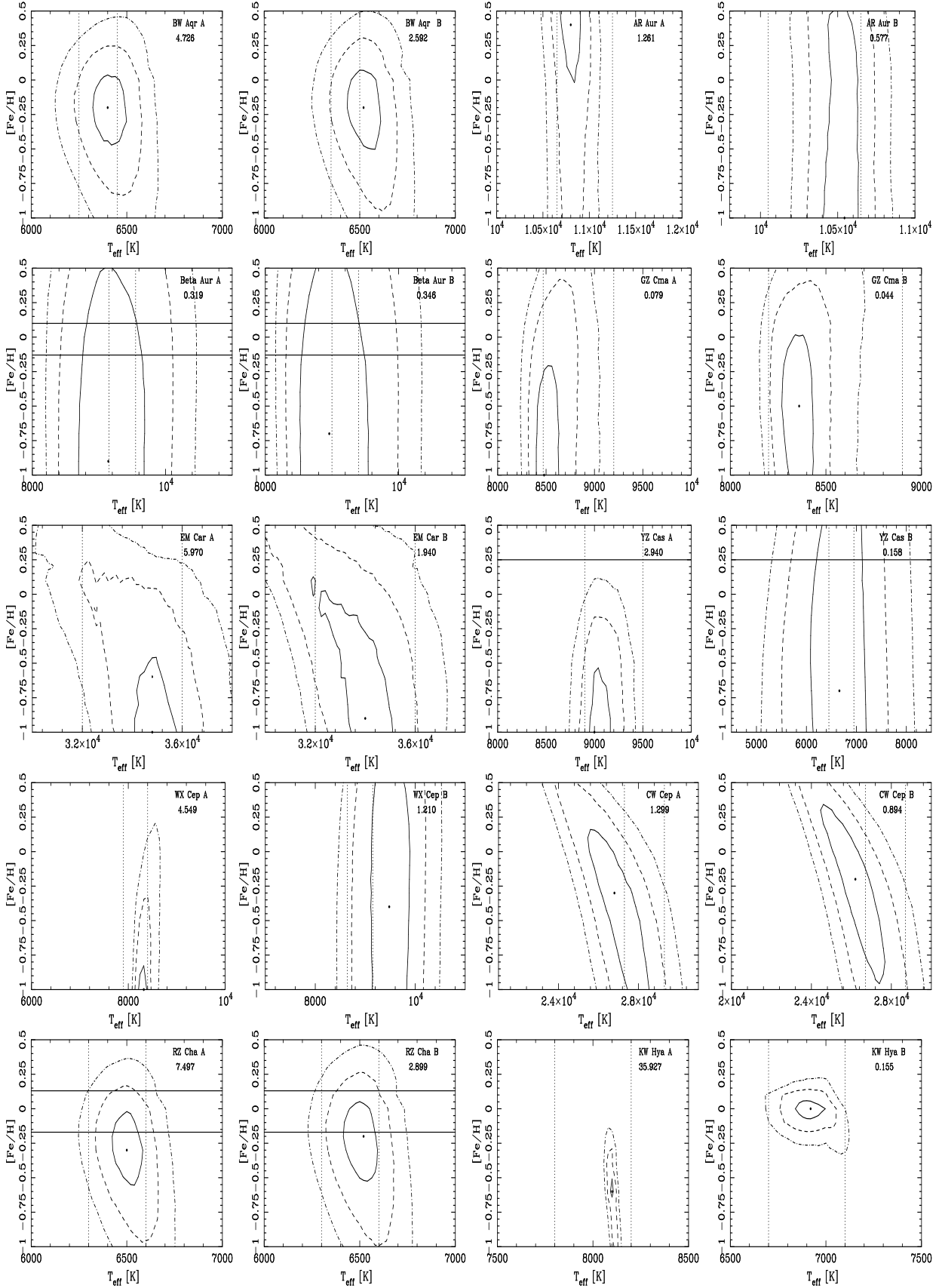

 \centerline{\psfig{file=8144.01.f4,width=4.truecm,height=4.5truecm,angle=-90.}
             \psfig{file=8144.02.f4,width=4.truecm,height=4.5truecm,angle=-90.}
             \psfig{file=8144.03.f4,width=4.truecm,height=4.5truecm,angle=-90.}
             \psfig{file=8144.04.f4,width=4.truecm,height=4.5truecm,angle=-90.}}
 \centerline{\psfig{file=8144.05.f4,width=4.truecm,height=4.5truecm,angle=-90.}
             \psfig{file=8144.06.f4,width=4.truecm,height=4.5truecm,angle=-90.}
             \psfig{file=8144.07.f4,width=4.truecm,height=4.5truecm,angle=-90.}
             \psfig{file=8144.08.f4,width=4.truecm,height=4.5truecm,angle=-90.}}
 \centerline{\psfig{file=8144.09.f4,width=4.truecm,height=4.5truecm,angle=-90.}
             \psfig{file=8144.10.f4,width=4.truecm,height=4.5truecm,angle=-90.}
             \psfig{file=8144.11.f4,width=4.truecm,height=4.5truecm,angle=-90.}
             \psfig{file=8144.12.f4,width=4.truecm,height=4.5truecm,angle=-90.}}
 \centerline{\psfig{file=8144.13.f4,width=4.truecm,height=4.5truecm,angle=-90.}
             \psfig{file=8144.14.f4,width=4.truecm,height=4.5truecm,angle=-90.}
             \psfig{file=8144.15.f4,width=4.truecm,height=4.5truecm,angle=-90.}
             \psfig{file=8144.16.f4,width=4.truecm,height=4.5truecm,angle=-90.}}
 \centerline{\psfig{file=8144.17.f4,width=4.truecm,height=4.5truecm,angle=-90.}
             \psfig{file=8144.18.f4,width=4.truecm,height=4.5truecm,angle=-90.}
             \psfig{file=8144.19.f4,width=4.truecm,height=4.5truecm,angle=-90.}
             \psfig{file=8144.20.f4,width=4.truecm,height=4.5truecm,angle=-90.}}

	\caption{Simultaneous solution of  {\teff} and {\feh} matching
	(b$-$y)$_0$, m$_0$, c$_0$, and {\lg}. The name of the star and
	the $\chi^2_{\mathrm  min}$  are   labelled directly  in  each
	panel.  When  available, effective  temperature determinations
	from previous studies ({\it vertical lines}) and observational
	indications of  metallicity ({\it horizontal  lines}) are also
	shown. }
\label{fig:all1}
\end{figure*}

\begin{figure*}[ht]
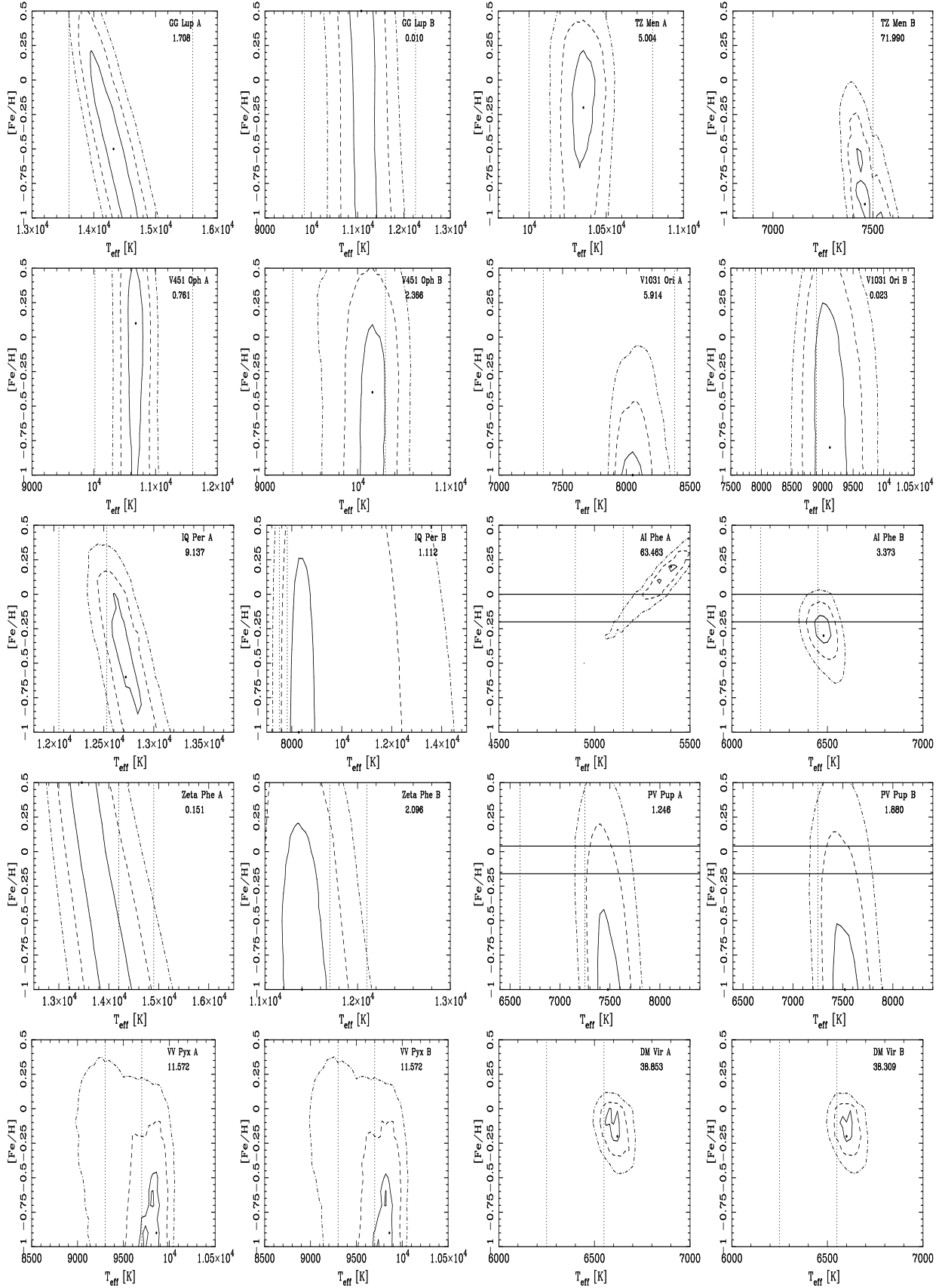

 \centerline{\psfig{file=8144.01.f5,width=4.truecm,height=4.5truecm,angle=-90.}
             \psfig{file=8144.02.f5,width=4.truecm,height=4.5truecm,angle=-90.}
             \psfig{file=8144.03.f5,width=4.truecm,height=4.5truecm,angle=-90.}
             \psfig{file=8144.04.f5,width=4.truecm,height=4.5truecm,angle=-90.}}
 \centerline{\psfig{file=8144.05.f5,width=4.truecm,height=4.5truecm,angle=-90.}
             \psfig{file=8144.06.f5,width=4.truecm,height=4.5truecm,angle=-90.}
             \psfig{file=8144.07.f5,width=4.truecm,height=4.5truecm,angle=-90.}
             \psfig{file=8144.08.f5,width=4.truecm,height=4.5truecm,angle=-90.}}
 \centerline{\psfig{file=8144.09.f5,width=4.truecm,height=4.5truecm,angle=-90.}   
             \psfig{file=8144.10.f5,width=4.truecm,height=4.5truecm,angle=-90.}
             \psfig{file=8144.11.f5,width=4.truecm,height=4.5truecm,angle=-90.}
             \psfig{file=8144.12.f5,width=4.truecm,height=4.5truecm,angle=-90.}}
 \centerline{\psfig{file=8144.13.f5,width=4.truecm,height=4.5truecm,angle=-90.}
             \psfig{file=8144.14.f5,width=4.truecm,height=4.5truecm,angle=-90.}
             \psfig{file=8144.15.f5,width=4.truecm,height=4.5truecm,angle=-90.}
             \psfig{file=8144.16.f5,width=4.truecm,height=4.5truecm,angle=-90.}}
 \centerline{\psfig{file=8144.17.f5,width=4.truecm,height=4.5truecm,angle=-90.}
             \psfig{file=8144.18.f5,width=4.truecm,height=4.5truecm,angle=-90.}
             \psfig{file=8144.19.f5,width=4.truecm,height=4.5truecm,angle=-90.}
             \psfig{file=8144.20.f5,width=4.truecm,height=4.5truecm,angle=-90.}}

	\caption{Same as Fig. \ref{fig:all1}.}
\label{fig:all2}
\end{figure*}

In Figures \ref{fig:all1} and  \ref{fig:all2} we show the full results
obtained (from {\by}, {\mun}, and  {\cun})  for all  the stars of  the
sample in ({\teff}, {\feh}) planes. All the ({\teff},{\feh})-solutions
inside   the contours allow   to   reproduce, at  different confidence
levels, both the   observed Str{\"o}mgren colours ({\by}, {\mun},  and
{\cun}) and  the surface gravity with the  BaSeL models.  As a general
trend,  it is  important to  notice  that  our  {\teff} ranges do  not
provide    estimates systematically  different   from   previous  ones
(vertical  dotted lines).    Furthermore,  the 3-$\sigma$   confidence
regions show that    most previous {\teff}  estimates  are optimistic,
except for some stars (e.g., GG  Lup A, TZ Men  A, and V451 Oph A) for
which our method gives  better constraints on the estimated  effective
temperature.  At a  1-$\sigma$  confidence level (68.3\%), our  method
often   provides better constraints  for  {\teff} determination. However, 
it is worth noticing  that for  a few  stars  the match  is really  bad (see
$\chi^{2}_{\rm min}$-values  labelled directly on Fig.  \ref{fig:all1}
and \ref{fig:all2}).  As already mentioned,  with 3 observational data
(\by,  m$_1$, c$_1$) and  2 free  parameters  ({\teff} and {\feh}), we
expect  to  find  a $\chi^2$-distribution   with 3$-$2$=$1  degree  of
freedom and a typical $\chi^{2}_{\rm min}$-value of about 1.  For some
stars (e.g.   VV Pyx, DM  Vir and  KW Hya A),  $\chi^{2}_{\rm min}$ is
greater than  10,  a  too  high value  to  be  acceptable  because the
probability to  obtain an observed  minimum $\chi$-square greater than
the value $\chi^{2}$$ =$10 is less than 0.2{\%}. For this reason, the 
results given for a particular star should not be used without carefully 
considering the $\chi^{2}_{\rm min}$-value. 

 One of the  most striking features appearing in  nearly all panels of
 Figs.
\ref{fig:all1} and \ref{fig:all2} is the considerable range of {\feh}
accepted inside the confidence levels.   This is particularly true for
stars hotter than $\sim$ 10,000 K (as, for instance, EM Car A \& B and
GG Lup A \&  B), for which  optical photometry is quite insensitive to
the stellar metal content.   For these stars, a  large range in {\feh}
gives very similar $\chi^2$ values. In contrast, for the coolest stars
in our sample,   our  method provides  straight constraints  on  their
metallicity. Actually, when observational metallicity indications are
available  ($\beta$  Aur, YZ Cas, RZ  Cha,  AI Phe,  and PV  Pup), the
contour solutions are found in good  agreement with previous estimated
{\feh} ranges (labelled as horizontal lines in Figs.
\ref{fig:all1} and \ref{fig:all2}).

The effective temperatures   derived from  our minimization  procedure
cannot  be easily  presented in a   simple  table format, as they  are
intrinsically related to metallicity.  We nonetheless provide in Table
3, as an indication of  the estimated stellar  parameters for all  the
stars  in our sample, the  best  ($\chi^2_{\mathrm min}$) simultaneous
solutions  ({\teff},{\feh}) for the   three following cases: by  using
{\by} and {\mun}   (Case 1), {\by} and {\cun}   (Case 2) and  by using
{\by}, {\mun}, and {\cun} (Case 3).   In Case 1 and  Case 2, a typical
$\chi^{2}_{\rm min}$-value close to   zero is theoretically  expected,
and   in  Case 3,  as previously  mentioned,   one  expects  a typical
$\chi^{2}_{\rm min}$-value of about 1. There are quite a few stars for
which $\chi^2_{\mathrm min}$ increases dramatically  between Case 1 or
2 and Case 3  to a clearly unacceptable value  (most notably AI  Phe A
between Case 1 and Case 3). This point means  that although a good fit
is     obtained   with  two   photometric    indices,   no  acceptable
$\chi^2_{\mathrm  min}$-value is obtained by adding  one more index in
Case 3. Consequently, Case 1 or Case 2-solutions have  to be chosen in
such  cases.  For  these stars,  even   if the  $\chi^2_{\mathrm min}$
solutions shown in   Figs.  \ref{fig:all1} and  \ref{fig:all2} are not
reliable, it  is interesting to notice  that  the contours derived are
however still in agreement with previous works.
 
The  surprising result in  Table 3  is  that  many solutions  are very
metal-poor.   This in fact    means  that  the  $\chi^{2}_{\rm   min}$
solutions are not {\em necessarily} the most realistic ones.  We must,
therefore,  emphasize that the values presented  in Table 3 should not
be  used without carefully  considering the  confidence level contours
shown  in Figs.  \ref{fig:all1} and \ref{fig:all2}.  For most stars in
our sample, {\teff} and {\feh} do not appear strongly correlated (i.e.
the confidence regions do not exhibit oblique shapes), but there are a
few cases for which the assumed metallicity leads to a different range
in the derived effective temperature (EM Car B, CW Cep A  \& B, GG Lup
A, $\zeta$  Phe   A).  These  results  point out  that   the classical
derivation of  {\teff} from  calibration without exploring  all {\feh}
values is not always a reliable method, even for hot stars.

\begin{table*}[h]
\begin{center}
	\caption[]{Best  simultaneous ({\teff},{\feh}) solutions using
	(b$-$y) and m$_1$ (Case 1), (b$-$y) and \\ c$_1$ (Case 2)  or  
        (b$-$y), m$_1$, and c$_1$ (Case 3).}
\begin{flushleft}
\begin{tabular}{lrrrrrrrrr}
\hline\noalign{\smallskip}
            &   \multicolumn{3}{c}{Case 1}  & \multicolumn{3}{c}{Case 2} &   \multicolumn{3}{c}{Case 3}  \\
\noalign{\smallskip}\hline\noalign{\smallskip}
Name        &   {\teff} & [Fe/H]   &  $\chi^{2}_{\rm min}$  &  {\teff} & [Fe/H] &  $\chi^{2}_{\rm min}$ 
            &   {\teff} & [Fe/H]   &  $\chi^{2}_{\rm min}$  \\
\noalign{\smallskip}\hline\noalign{\smallskip}
BW Aqr      &  6220 & -0.5 & 0.01 &    6400  &    0.5 &  0.75 &    6400  &   -0.2 &  4.73 \\
            &  6400 & -0.4 & 0.00 &    6540  &    0.5 &  0.27 &    6520  &   -0.2 &  2.59 \\
AR Aur      & 11240 &  0.5 & 0.00 &   10760  &    0.5 &  0.66 &   10800  &    0.4 &  1.26  \\
            & 10352 & -0.2 & 0.00 &   10568  &   -0.6 &  0.00 &   10544  &   -1.0 &  0.58 \\
$\beta$ Aur &  9260 & -0.8 & 0.13 &    9140  &   -0.9 &  0.18 &    9140  &   -0.9 &  0.32 \\
            &  8900 &  0.2 & 0.00 &    9020  &   -1.0 &  0.19 &    8960  &   -0.7 &  0.35 \\
GZ Cma      &  8480 & -1.0 & 0.06 &    8480  &   -0.9 &  0.02 &    8480  &   -1.0 &  0.08  \\
            &  8340 & -0.6 & 0.00 &    8380  &   -1.0 &  0.01 &    8360  &   -0.5 &  0.04   \\
EM Car      & 30640 & -0.9 & 2.72 &   35920  &    0.5 &  0.01 &   34800  &   -1.0 &  5.97  \\
            & 30000 & -1.0 & 0.50 &   32240  &    0.2 &  0.01 &   34000  &   -0.9 &  1.94 \\
YZ Cas      &  9080 & -1.0 & 0.00 &    9000  &   -1.0 &  2.93 &    9000  &   -1.0 &  2.94 \\
            &  6820 & -0.2 & 0.00 &    6660  &   -1.0 &  0.08 &    6660  &   -0.7 &  0.16 \\
WX Cep      &  8380 & -1.0 & 1.57 &    8180  &    0.2 &  0.00 &    8280  &   -1.0 &  4.55  \\
            &  9960 &  0.4 & 0.00 &    9480  &   -0.2 &  1.10 &    9480  &   -0.4 &  1.21  \\
CW Cep      & 31000 &  0.4 & 0.00 &   26800  &   -0.3 &  0.01 &   26800  &   -0.3 &  1.30  \\
            & 29400 &  0.5 & 0.25 &   24000  &    0.5 &  0.00 &   26200  &   -0.2 &  0.89  \\
RZ Cha      &  6240 & -0.6 & 0.02 &    6500  &    0.5 &  2.06 &    6500  &   -0.3 &  7.50 \\
            &  6340 & -0.4 & 0.01 &    6520  &    0.5 &  0.40 &    6520  &   -0.2 &  2.90 \\
KW Hya      &  7860 & -0.8 & 0.06 &    8120  &   -1.0 & 34.60 &    8100  &   -0.6 & 35.93 \\
            &  6900 &  0.0 & 0.03 &    6940  &    0.4 &  0.02 &    6920  &    0.0 &  0.15  \\
GG Lup      & 14320 & -0.5 & 1.70 &   14320  &   -0.5 &  0.08 &   14320  &   -0.5 &  1.71 \\
            & 11000 &  0.5 & 0.00 &   11080  &    0.5 &  0.00 &   11080  &    0.5 &  0.01   \\
TZ Men      & 10780 & -0.4 & 0.00 &   10255  &    0.2 &  0.01 &   10352  &   -0.2 &  5.00  \\
            &  7220 & -1.0 & 1.51 &    7440  &    0.4 & 43.51 &    7460  &   -0.9 & 71.99  \\
V451 Oph    & 11160 & -0.4 & 0.05 &   10620  &    0.3 &  0.05 &   10680  &    0.1 &  0.76  \\
            & 10480 & -1.0 & 0.63 &   10080  &   -0.8 &  0.25 &   10160  &   -0.4 &  2.37  \\
V1031 Ori   &  8080 & -1.0 & 5.62 &    7990  &   -1.0 &  0.06 &    8050  &   -1.0 &  5.91 \\
            &  9120 & -0.8 & 0.02 &    9120  &   -0.8 &  0.00 &    9120  &   -0.8 &  0.02   \\
IQ Per      & 13600 & -0.9 & 7.62 &   12760  &   -0.7 &  0.10 &   12720  &   -0.6 &  9.14 \\
            &  8600 & -1.0 & 0.88 &    8120  &    0.5 &  0.09 &    8280  &   -1.0 &  1.11  \\
AI Phe      &  4860 & -0.9 & 0.11 &    4860  &   -0.1 & 42.05 &    5400  &    0.2 & 63.46 \\
            &  6360 & -0.3 & 0.03 &    6420  &    0.3 &  0.01 &    6480  &   -0.3 &  3.37   \\
$\zeta$ Phe & 12820 &  0.5 & 0.00 &   13620  &    0.2 &  0.01 &   13460  &    0.5 &  0.15  \\
            & 11000 & -1.0 & 1.50 &   11400  &   -1.0 &  1.34 &   11400  &   -1.0 &  2.10 \\
PV Pup      &  7440 & -1.0 & 1.21 &    7520  &   -1.0 &  0.09 &    7480  &   -1.0 &  1.25 \\
            &  7440 & -1.0 & 1.71 &    7520  &   -0.5 &  0.01 &    7520  &   -1.0 &  1.88 \\
VV Pyx      &  9260 & -0.9 & 8.55 &    9020  &    0.5 &  0.18 &    9860  &   -0.9 & 11.57   \\
            &  9260 & -0.9 & 8.55 &    9020  &    0.5 &  0.18 &    9860  &   -0.9 & 11.57   \\
DM Vir      &  6360 & -0.3 & 0.15 &    6600  &    0.5 & 15.41 &    6620  &   -0.2 & 38.85  \\
            &  6360 & -0.3 & 0.15 &    6600  &    0.5 & 15.41 &    6600  &   -0.2 & 38.31 \\
\noalign{\smallskip}
\hline
\end{tabular}
\end{flushleft}
\end{center}
\label{tab:res}
\end{table*}

\subsection{Comparison with Hipparcos parallax}

\begin{figure*}[ht]

\centerline{\psfig{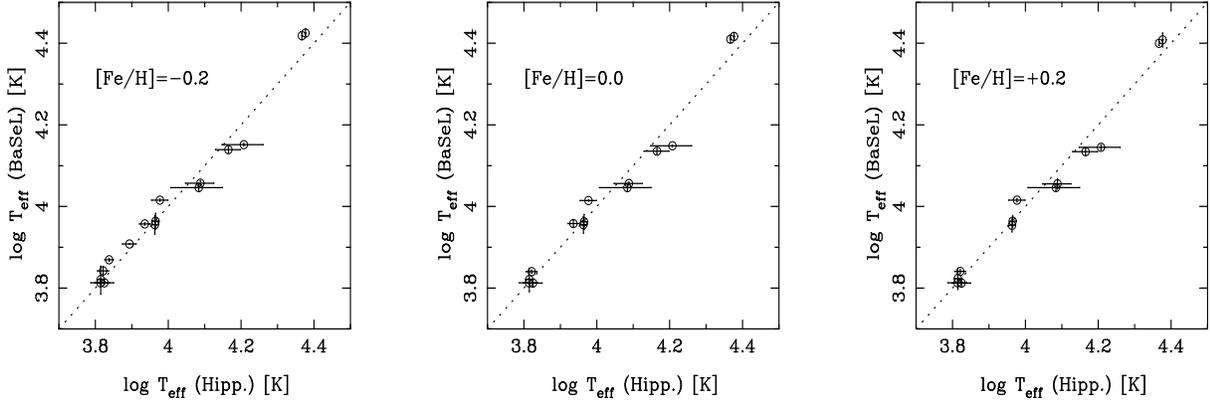}}
	\caption{Hipparcos-       versus      BaSeL-derived  effective
	temperatures for $\beta$ Aur, YZ Cas, CW  Cep, RZ Cha, KW Hya,
	GG Lup, TZ Men, and $\zeta$ Phe.  The  errors are not shown on
	the Hipparcos axis for CW  Cephei (the hottest binary in these
	figures).  See text for explanation.}
\label{fig:hipp}
\end{figure*}

Very recently, Ribas   {\al}   (1998)  have computed   the   effective
temperatures of   19  eclipsing binaries   included  in the  Hipparcos
catalogue  from  their radii,  Hipparcos trigonometric parallaxes, and
apparent visual  magnitudes   corrected  for absorption.   They   used
Flower's (1996) calibration to derive  bolometric corrections.  Only 8
systems are in common with our working sample. The comparison with our
results is made in Table 4. 
The {\teff} being highly related with metallicity, a direct comparison
is  not  possible because,   unlike  the Hipparcos-derived   data, our
results are not given in terms of temperatures with error bars, but as
ranges of {\teff} compatible  with  a given  {\feh}. Thus, the  ranges
reported in Tab. 4 
are given assuming  three different hypotheses: {\feh}$=-$0.2,  {\feh}
$=$    0,   and {\feh}   $=$  0.2.    The  overall  agreement is quite
satisfactory, as    illustrated  in  Fig.    \ref{fig:hipp}.   \\  The
disagreement for the temperatures of CW Cephei can be explained by the
large error  of  the  Hipparcos parallax ($\sigma$$_{\rm   \pi}$/$\pi$
$\simeq$70\%).  For  such  large  errors, the Lutz-Kelker   correction
(Lutz \&  Kelker 1973) cannot   be neglected: the average  distance is
certainly underestimated and,  as a consequence,  the  {\teff} is also
underestimated in   Ribas   {\al}'s  (1998) calculation.    Thus,  the
agreement with the results obtained from the BaSeL models is certainly
better than it   would appear in   Fig.  \ref{fig:hipp} and   Tab.  4.
Similar corrections, of   slightly lesser  extent, are  probably  also
indicated   for  the {\teff}  of  RZ   Cha  and  GG   Lup,  which have
$\sigma$$_{\rm
\pi}$/$\pi  >$ 10{\%} (11.6\% and  11.4\%, respectively).  Finally, it
is worth noting  that the system with the  smallest relative  error in
Tab.    4, $\beta$  Aur,  shows  excellent  agreement between  {\teff}
(Hipparcos)  and {\teff}   (BaSeL), which underlines  the  validity of 
the BaSeL models.

\section{Conclusion}

The comprehensive knowledge  of fundamental parameters of single stars
is  the basis of the  modelling of star clusters   and galaxies.  Most
fundamental  stellar parameters of  the  individual components in  SB2
eclipsing binaries are known  with very high accuracy.  Unfortunately,
while masses and radii are well  determined, the temperatures strongly
depend on  photometric calibrations.  In this  paper, we  have used an
empirically-calibrated  grid of   theoretical  stellar  spectra (BaSeL
models) for simultaneously deriving homogeneous effective temperatures
and metallicities  from observed data.  Although  a few stars  show an
incompatibility between the observed  and synthetic $uvby$ colours  if
we try to match the three Str{\"o}mgren indices (\by),
\mun, and \cun, the  overall determinations are  satisfying.  Moreover,
an acceptable solution  is always possible  when only  considering two
photometric indices, as in Case 1 or Case  2 (see Table 3).  The large
range of {\feh} associated  with acceptable confidence levels makes it
evident   that   the  classical    method   to  derive    {\teff} from
metallicity-independent    calibrations   should  be considered   with
caution.    We found that,  even  for  hot stars  for  which we expect
optical   photometry   to  be  nearly     insensitive to  the  stellar
metal-content, a  change  in  the  assumed metallicity  can  lead to a
significant   change  in  the  predicted effective  temperature range.
Furthermore, for cool stars, both {\teff} and  {\feh} can be estimated
with good  accuracy from   the photometric  method.   The   effects of
surface gravity  and  interstellar reddening have also  been carefully
studied.  In  particular, an apparently minor  error in  reddening can
change dramatically  the shape of  the confidence contour levels, and,
therefore, the parameter values hence  derived.  By exploring the best
$\chi^2$-fits  to the   photometric   data,  we  have  re-derived  new
reddening values for  some stars (see  Table 1).  Finally, comparisons
for  16  stars with  Hipparcos-based {\teff}  determinations show good
agreement with the temperatures derived  from  the BaSeL models.   The
agreement  is even  excellent for  the star having  the  most reliable
Hipparcos data in the sample studied  in this paper. These comparisons
also   demonstrate  that, while originally     calibrated in order  to
reproduce the broad-band  (UBVRIJHKL) colours,  the BaSeL models  also
provide   reliable results   for medium-band photometry   such  as the
Str{\"o}mgren photometry.  This   point gives a significant weight  to
the    validity of the    BaSeL   library  for synthetic    photometry
applications in general.

\begin{table*}[h]
\begin{center}
	\caption[]{Effective temperatures  from Hipparcos (after Ribas
	{\al} 1998) and from BaSeL models \\matching (\by)$_0$, m$_0$,
	c$_0$,  and {\lg}    for the  three   following metallicities:
	{\feh}$=-$0.2, {\feh} $=$ 0
\\and {\feh} $=$ 0.2.}
\begin{flushleft}
\begin{tabular}{lrrcrcrc}
\hline\noalign{\smallskip}
Name         &       &      \multicolumn{2}{c}{{\feh}$=-$0.2}        &
\multicolumn{2}{c}{{\feh} $=$ 0.} & \multicolumn{2}{c}{{\feh} $=$ 0.2} \\
\noalign{\smallskip}\hline\noalign{\smallskip}
     &  {\teff}(Hipp.)  [K]   &   {\teff}(BaSeL) [K]   &  $\sigma$  &
     {\teff}(BaSeL) [K] &  $\sigma$ & {\teff}(BaSeL) [K] & $\sigma$
     \\
\noalign{\smallskip}
\hline\noalign{\smallskip}
$\beta$ Aur & 9230$\pm$150   & [8780,9620]   & 1 & [8780,9560]   & 1 & [8900,9500]   & 1 \\
            & 9186$\pm$145   & [8540,9500]   & 1 & [8600,9440]   & 1 & [8660,9320]   & 1 \\
YZ Cas      & 8624$\pm$290   & [9000,9120]   & 2 & [8920,9240]   & 3 & no solution   &   \\
            & 6528$\pm$155   & [6100,7140]   & 1 & [6180,7060]   & 1 & [6260,7060]   & 1 \\
CW Cep      & 23804          & [26000,27200] & 1 & [25600,26600] & 1 & [24600,26600] & 2 \\
            & 23272          & [25600,26800] & 1 & [25200 26200] & 1 & [24800,25400] & 1 \\
RZ Cha      & 6681$\pm$400   & [6440,6560]   & 1 & [6380,6600]   & 2 & [6340,6640]   & 3 \\ 
            & 6513$\pm$385   & [6420,6580]   & 1 & [6460,6540]   & 1 & [6420,6580]   & 2 \\
KW Hya      & 7826$\pm$340   & [8080,8100]   & 3 & no solution   &   & no solution   &   \\
            & 6626$\pm$230   & [6780,7120]   & 3 & [6860,6980]   & 1 & [6860,7000]   & 3 \\
GG Lup      & 16128$\pm$2080 & [14080,14260] & 1 & [14020,14140] & 1 & [13780,14140] & 2 \\
            & 12129$\pm$1960 & [10920,11320] & 1 & [10920,11320] & 1 & [10920,11320] & 1 \\
TZ Men      & 9489$\pm$490   & [10300,10420] & 1 & [10300,10380] & 1 & [10260,10460] & 2 \\
            & 6880$\pm$190   & [7340,7460]   & 3 & no solution   &   & no solution   &   \\
$\zeta$ Phe & 14631$\pm$1150 & [13540,14020] & 1 & [13460,13860] & 1 & [13380,13860] & 1 \\
            & 12249$\pm$1100 & [11240,11560] & 1 & [11280,11480] & 1 & [11040,11680] & 2 \\
\noalign{\smallskip}
\hline
\end{tabular}
\end{flushleft}
\end{center}
\label{tab:Hipp}
\end{table*}

\begin{acknowledgements}
  E.  L. gratefully  thanks the Swiss  National Science Foundation for
  financial  support and, in particular, Professor   R.  Buser and the
  Astronomisches   Institut   der  Universit\"at  Basel    for   their
  hospitality.  We   acknowledge  the referee,  Dr   Pols, for helpful
  comments  which have  improved  the  clarity  of  this paper.   This
  research has made   use  of the  Simbad  database  operated at  CDS,
  Strasbourg, France, and was supported  by the Swiss National Science
  Foundation.

\end{acknowledgements}

\end{document}